\documentclass[apj]{emulateapj}
\usepackage{amssymb}
\usepackage{multirow}
\usepackage{graphicx}
\usepackage{subfigure}

\shorttitle{DLAs and the Smallest Galaxies}
\shortauthors{Webster, Sutherland \& Bland-Hawthorn}

\newcommand{\comments}[1]{}

\begin{document}

\title{The Chemical Evolution of Very Metal-Poor Damped Lyman-$\alpha$ Systems}

\author{David Webster}
\affil{Sydney Institute for Astronomy, School of Physics, University of Sydney, NSW 2006, Australia}

\author{Joss Bland-Hawthorn}
\affil{Sydney Institute for Astronomy, School of Physics, University of Sydney, NSW 2006, Australia}
\email{d.webster@physics.usyd.edu.au}

\author{Ralph S. Sutherland}
\affil{Research School of Astronomy \& Astrophysics, Australian National University, Cotter Rd,
Weston, ACT 2611, Australia}

\begin{abstract}

In earlier work we showed that a dark matter halo with a virial mass of $10^7$~M$_\odot$ can survive feedback from its own massive stars and form 
stars for $\gtrsim100$~Myr. We also found that our modelled systems were consistent with observations of ultrafaint dwarfs (UFDs), the least massive 
known galaxies. Very metal-poor damped Lyman-$\alpha$ systems (DLAs) recently identified at $z\sim2$ may represent the gas that formed at least some 
of the observed stars in UFDs. We compare projected sightlines from our simulations to the observed 
metal-poor DLAs and find that our models can reach the densities of the observed sightlines; however the metallicities are 
inconsistent with the single supernova simulations, suggesting enrichment by multiple supernovae. 
We model two scenarios for the history of these systems. The first
 explains the gas abundances in DLAs by a single burst of star formation. This model can produce the observed DLA abundances, but does not provide 
an explanation as to why the DLAs show suppressed [$\alpha$/Fe] compared to the stellar population of UFDs. The second scenario splits 
the DLAs into a population which is enriched by a single burst, and a population that is enriched by a second burst after the accretion of 
metal-poor gas. In this scenario, the suppressed 
average [$\alpha$/Fe] in DLAs compared to UFDs results from enrichment of second-burst systems by Type Ia supernovae.

\end{abstract}


\section{Introduction}
%Need to add discussion/justification of relationship between DLAs and UFDs.

The very metal-poor Damped Lyman-$\alpha$ systems (DLAs) with [Fe/H]~$<-2$ recently discovered by \citet{cooke11a,cooke11b,cooke13,cooke14} have chemical signatures 
suggesting that they have experienced only a few enrichment events, making them promising candidates for probing the chemical signatures of the first generations of stars. DLAs absorb 
light from a bright background source, allowing the detection of gas that is too faint to observe in emission.  The most distant and metal-poor DLAs may be the environments of the formation of some of the first generations of stars in the universe. 
Along with cosmological hydrodynamical simulations \citep{pontzen08,fumagalli11,cen12,bird13}, the few DLAs at $z > 2$ that have been detected in emission \citep{krogager12,jorgenson14} 
suggest that the host galaxies of higher metallicity ($Z\sim1/50$) DLAs are mostly 
$10^9-10^{11}$~M$_\odot$ dwarf galaxies. 

\citet{cooke14} found that the kinematics of the very metal-poor DLAs agreed with 
those of Milky Way dwarf spheroidal galaxies. Furthermore, the decline in [$\alpha$/Fe] for [Fe/H] $\gtrsim$ -2.0, which is usually assumed to indicate 
enrichment from Type Ia supernovae \citep{tinsley79}, is similar to that observed in the dwarf 
spheroidals, indicative of a similar star formation history. This differs from the Milky Way halo, 
which shows approximately constant [$\alpha$/Fe] 
for $-3.75<~$[Fe/H]$~<-0.75$. \citet{cooke14} therefore suggest that the very metal-poor DLAs correspond to the least massive systems that can form 
stars at $z\sim3$.

Over the past decade, a number of galaxies with luminosities below $10^5$~L$_\odot$ have been discovered in the local universe. Known as ultrafaint 
dwarfs (UFDs), these systems are promising candidates for probing the chemical signatures of the first generations of stars \citep{bovill09,frebel12,vargas13,frebel14}. 
The majority of UFDs contain only ancient stars, with the spread of ages being $<2$~Gyr 
\citep{munoz10, brown12, vargas13}. \citet{brown14} found that the stellar populations of five UFDs: 
Hercules, Leo IV, Ursa Major I, Bootes 1 and Coma Berenices showed similar ages, suggesting that a global event such as reionization truncated 
their star formation. \citet{weisz14b} found that most galaxies with $M_* < 10^5$~M$_\odot$ formed $\gtrsim80\%$ of their 
stars before $z\sim2$, although there was significant variation between individual galaxies. 
For example, Canes Venatici II formed stars until $\sim8-11~$Gyr ago, while Hercules and Leo IV 
formed 90~\% of their stellar mass by 11-12~Gyr. The star formation history of UFDs indicates that at least some of them were forming stars at $z\sim3$ and therefore must have 
contained neutral gas at this time. This suggests the possibility that very metal-poor DLAs trace gas in UFDs at the time they were forming stars. 

%UFDs such as Segue 1 may have formed only $\sim$1000 stars and experienced only a single star forming burst 
%\citep{frebel14}, meaning that they were likely enriched by only a few Type II supernovae and no Type Ia supernovae. Other UFDs had a more extended 
%star-forming history, however still experienced few enrichment events, such that the signatures of the first generations of stars are likely to remain relatively intact.

With current instruments, it is difficult to observe UFDs beyond the local Milky Way + M31 system, although there has been one detection in the 
Virgo cluster \citep{jang14}. The relationship between low mass fossil galaxies that formed in the early universe and the observed UFDs today is therefore complicated 
by the influence of environment. Current observations cannot distinguish between systems which are true fossils of reionization and those that had their 
star formation quenched by other processes, such as tidal stripping \citep{weisz14a}. The UFDs that may be traced by DLAs need not be close to a host 
galaxy and could provide an opportunity to study the smallest galaxies in a less complicated environment.

\citet{weisz14b} studied 13 galaxies identified by \citet{bovill11} as fossil candidates, 
finding that Hercules and Leo IV were the only strong candidates based on their star formation history. The Cooke DLAs trace gas at $z\sim2-4$, 
indicating that the systems they trace were not permanently quenched by reionization. This is consistent with the 
\citet{weisz14b} star formation histories for systems such as Canes Venatici II, for which the best fit model formed the bulk of its stars at 
$z\sim2$. The mass of Canes Venatici II is $1.4\times10^6$~M$_\odot$ within the half-light radius \citep{wolf10}, suggesting that even very low-mass 
systems can form stars at the redshifts of the Cooke DLAs.

In this work we use the hydrodynamical simulations of a $M_{\rm{vir}} = 10^7$~M$_\odot$ galaxy from \citet{hawthorn15}, in conjunction with the 
chemical evolution model from \citet{webster14}, to investigate possible scenarios for the history and evolution of DLAs. Two main processes prevent galaxies with lower masses from forming or surviving in the early universe. The first is the energy output of massive 
stars. In previous work \citep{hawthorn11, hawthorn15,webster14}, 
we presented hydrodynamical simulations of galaxies with formation masses of $10^7$~M$_\odot$ and lower, showing that the limit for stars 
to retain gas and form stars in the face of feedback from their own star formation is $M_{\rm{vir}}\approx10^{6.5}-10^7$~M$_\odot$. 
Furthermore, our models were able to reproduce most features of the observed 
relationship between [$\alpha$/Fe] and [Fe/H] in ultrafaint dwarf galaxies. 

The second process that can remove neutral gas from galaxies is the epoch of reionization, when the radiation from the first stars photoionised much of 
the neutral gas in the universe. It is often claimed \citep{rees86, barkana99, gnedin00, okamoto08} that neutral gas is unable to 
cool onto halos with masses $\lesssim10^8$M$_\odot$. However, more recent simulations \citep{bovill09} 
suggest that this is not a hard limit, with the influence of environment meaning that at least some lower mass systems can survive the epoch 
of reionization. \citet{ricotti05} used cosmological simulations to predict the existence of an undetected population of ultrafaint galaxies, 
around the same time as the first UFDs were observed. 

We now consider whether $10^7$~M$_\odot$ halos can explain the number of observed very metal-poor DLAs. 
For each $10^{11}~$M$_\odot$ halo, there are $\sim10^4$ halos with 
$M\leq10^7$~M$_\odot$. At $z\sim2-4$, the virial radius of a $10^{11}$~M$_\odot$ halo was $\sim120~$kpc \citep[calculated from][]{bryan98}, while the virial radius of a $10^7$~M$_\odot$ halo was $\sim0.5$~kpc (see Figure 1 and \citet{hawthorn15}). The ratio of cross-sectional area is proportional to the ratio of their radii squared, which is 
$\sim2\times10^{-5}$. However, $10^7$~M$_\odot$ halos are $10^4$ times more common, so the total cross-sectional area is $\sim20\%$ that of the $10^{11}$~M$_\odot$ 
halo. While this is only an approximate order of magnitude argument, it suggests that DLAs from $10^7$~M$_\odot$ halos should be frequent enough to be observed.

In Section~\ref{s:simulations}, we will summarise the simulations, although we refer to our previous work for the details. Section~\ref{s:comparison} 
compares the Cooke DLAs to our models. Possible scenarios for the history of the DLAs and their link to present-day UFDs are discussed in 
Section~\ref{s:scenarios}, followed by our conclusions in Section~\ref{s:conclusions}.
 
\section{Simulations}
\label{s:simulations}

The simulations used in this work are described in detail in \citet{hawthorn15} \& \citet{webster14}. 
In \citet{hawthorn15}, we presented high-resolution simulations performed using the hydro/ionization code \emph{Fyris Alpha} showing the effect of a 25~M$_\odot$ star 
on low-mass dark matter halos. An ionised region is created around the star prior to the supernova, resulting in the supernova having a much greater 
effect than for a lower mass star. The thermal and ionisation structures were calculated with the MAPPINGS IV ionisation code, with the ATLAS9 atmospheric grid \citep{castelli04} and the \citet{meynet02} evolutionary tracks used to model the progenitor star. The full details can be found in Section 2 of \citet{webster14}.

As a check for convergence, we investigated the amount of energy lost given a single-level resolution model, a two-level model with three times more cells in the central region, and a three-level model with the resolution again higher by a factor of 3. If the resolution is insufficient, 
the gas has a tendency to overcool. The three models showed only minor differences, and we 
therefore used the two-level model with the higher resolution level of 216$^3$ cells having a resolution of $\sim2~$pc/cell. 
The other possible effect of insufficient resolution is that the metals will mix too efficiently. 
 
The inclusion of radiative cooling and an inhomogeneous interstellar medium reduces the coupling between 
the supernova energy and the gas, such that halos with masses 
$M_{\rm{vir}}=10^{6.5-7}$~M$_\odot$ retain a large proportion of their baryons in the face of the ionization and supernova of the 25~M$_\odot$ star. 
The location of the star in the halo is also important, with supernovae occurring away from the centre resulting in lower levels of enrichment and 
greater retention of dense gas.
 
The simulations start with an intergalactic medium enriched to [Fe/H]~$=-4$ by the first stars. Such low [Fe/H] has been observed in the IGM at $z\sim3-3.5$ \citep{fumagalli11}. Our enhanced C, O and $\alpha$ elements ([$\alpha$/Fe]$_{\rm{init}} = 0.7$) give $Z_{\rm{init}} = 10^{-3.2}~$Z$_\odot$. We take our initial metallicity to be the critical threshold for low-mass star formation, such that low mass star formation proceeds only after the initial 25~M$_\odot$ star enriches the gas. While our threshold fits the critical metallicity $Z_{\rm{crit}}\sim10^{-4}-10^{-3}$~Z$_\odot$ of \citet{smith07}, other authors 
have suggested that this threshold could be as low at $Z_{\rm{crit}}\sim10^{-6}~$Z$_\odot$ \citep{schneider06}. As discussed in \citet{webster14}, removing the assumption that low-mass stars do not form at our starting metallicity does not have a large impact on the results, as it affects 
only the first 10~Myr of the 600~Myr star formation history. 

The gas mass within the scale radius $r_s$ of the Einasto dark matter potential is set to be 10\% of the dark matter mass within this radius. 
The overall baryon fractions $M_{\rm{gas,vir}}/M_{\rm{vir}}$ are $\approx12\%$. The model of an $M_{\rm{vir}} = 10^7$~M$_\odot$ halo 
contains 2.34$\times10^5$~M$_\odot$ of gas 
within $r_{\rm{s}}$ and 
1.5$\times10^6$~M$_\odot$ within $r_{\rm{vir}}$.

In the hydrodynamical simulations presented in \citet{hawthorn15}, the evolution of the gas after a single supernova was traced for 60~Myr. 
Longer time periods were not possible in these high-resolution simulations due to the effect of boundary conditions, while difficulties involved with studying the interaction between supernovae restricted the models to a single supernova. In \citet{webster14}, the density and metallicity of the gas 
after a single supernova was used as a template for the condition of the gas after subsequent supernovae. This allowed the density and metallicity of the 
gas to be tracked for 600~Myr. We model our halos in isolation, so do not include infall of diluting gas from the IGM, although we do consider the effects 
of dilution in Section 4.2, where the second burst of stars forms after the accretion of metal-poor gas.

The hydrodynamical simulation does not trace molecular cooling as would be required for the gas to cool to star formation temperatures. 
Instead, stars are allowed to form in the gas using the method of \citet{argast00}, in which 
$10^4$ cells are randomly selected, with each given a probability of forming a star proportional to the square of the density of the gas in the cell. The 
masses of the stars formed were selected by sampling a \citet{kroupa01} initial mass function. After each supernova, the density distribution of the 
gas is reset to the distribution just after the supernova in the single-supernova hydrodynamical simulation and the density then evolves as 
for the hydrodynamical simulation. The metallicity is treated in the same way, except that it is added to rather than reset, such that the enrichment from each supernova adds to the enrichment from all previous supernovae.

The yields used to determine [Fe/H] \& [$\alpha$/Fe] are from \citet{woosley95}, interpolated and extended to 8~M$_\odot$, taken to be the lowest mass star that ends its life as a supernova. After 100~Myr, Type Ia 
supernovae occur with a probability equivalent to the Type Ia supernovae rate determined by \citet{jimenez14}, scaled to the star formation rates of our systems. The Type Ia yields 
are from \citet{iwamoto99}.

\begin{figure}%[htp]
     \centering
     \includegraphics[width=.45\textwidth]{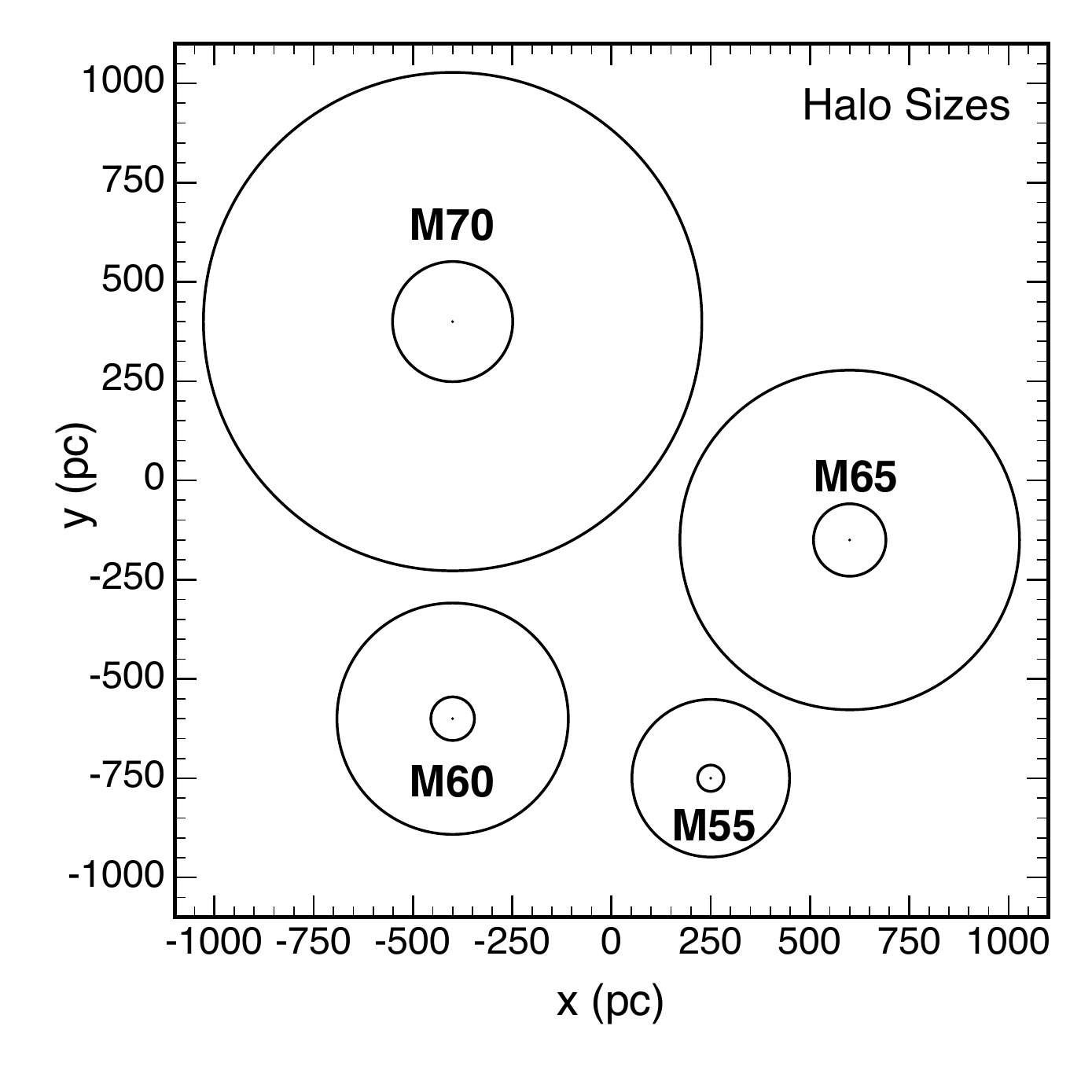}
     \caption{The physical sizes of the modelled halos at $z=10$. M55 denotes a dark matter virial mass of $10^{5.5}$~M$_\odot$, M60 has $M_{\rm{vir}}=10^6~$M$_\odot$ etc. The inner circles represent the scale radii, while the outer circles are the virial radii. For the off-centred explosion, the supernova is placed at 
the radius enclosing half the gas inside the scale radius, which is approximately halfway between the centre and the scale radius. See \citet{hawthorn15} 
for the full set of parameters.}
     \label{f:halosize}           
\end{figure}

\begin{figure}[htp]             
      \includegraphics[width=.45\textwidth]{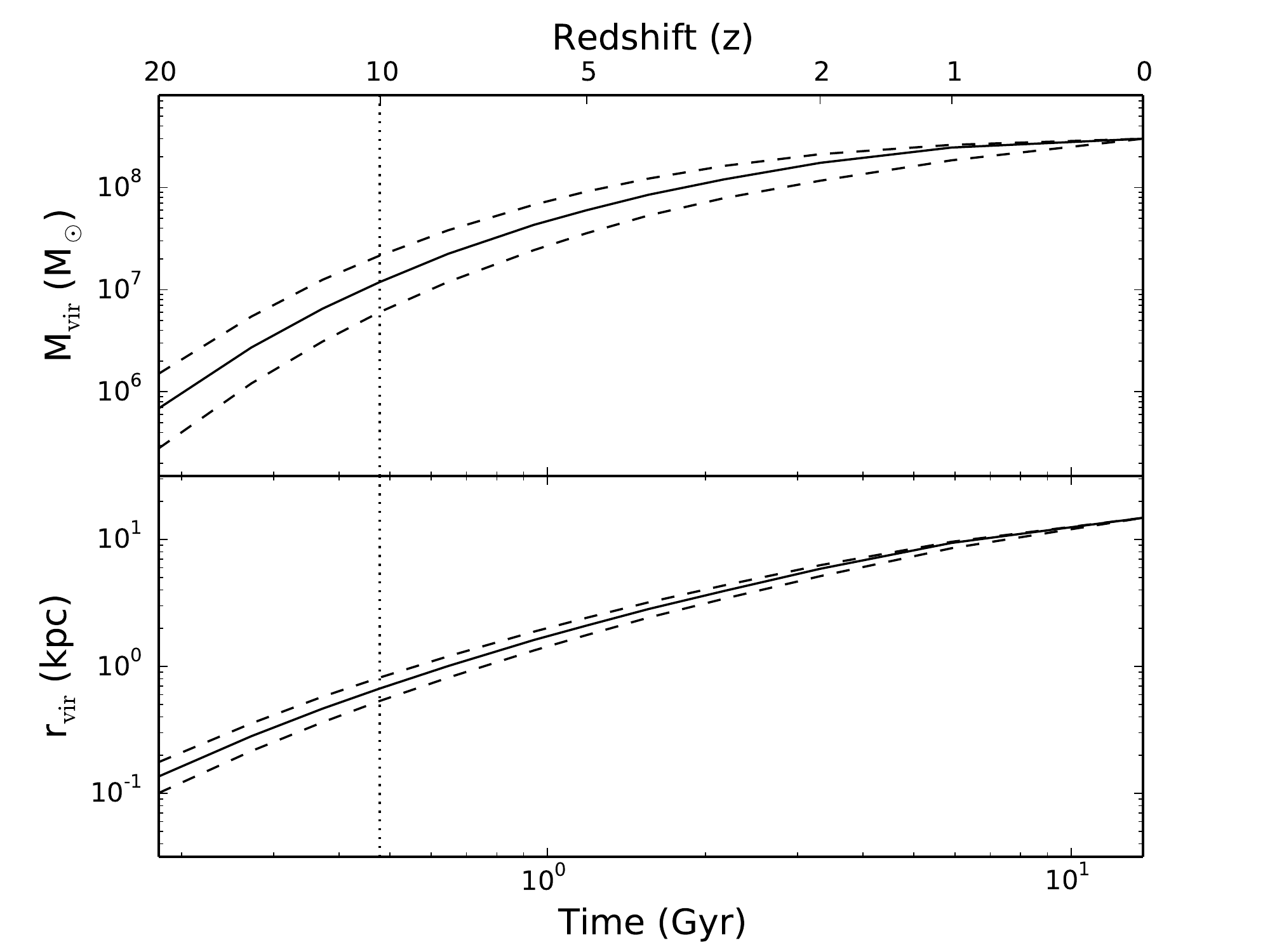}
      \caption{Evolution of the virial mass and radius for a halo with $M_{\rm{vir}}=10^7$~M$_\odot$ halo at $z=10$ (indicated by the dotted line) from $z=20$ to the present time. This plot was produced using data from 5000 runs of the tree merger code of \citet{parkinson08}. The dashed lines show the range which 67\% of 
halos fall within.}
\label{f:haloevol}
\end{figure}

Fig.~\ref{f:halosize} shows the halos used in this work. The $M_{\rm{vir}} = 10^{5.5}$~M$_\odot$, $10^{6}$~M$_\odot$, $10^{6.5}$~M$_\odot$ and 
$10^{7}$~M$_\odot$ models are referred to as M55, M60, M65 and M70 respectively. The scale radius, which contains nearly all the gas sufficiently dense 
to be observed as DLAs, ranges from 33~pc for M55 to 151~pc for M70, while the virial radii range from 200~pc for M55 to 630~pc for M70. These halo properties are at $z=10$. As shown in Figure~\ref{f:haloevol}, halos with $M_{\rm{vir}}=10^{6-7}$ at $z=10$ grow by an order of magnitude in both mass and radius by $z=3$. Their virial radii grow from 0.3-0.6~kpc to 1-3~kpc. The increase 
in halo mass will improve the gas retention compared to that of the first supernova. Our assumption that the first supernova occurs at $z=10$ is 
supported by \citet{power14}, which shows that halos with present-day masses $M_{\rm{vir}}<10^9~M_\odot$ 
reach the atomic cooling threshold at approximately this redshift. 

The suffixes we use in the rest of this work represent the type of model: CCH represents a model with a central explosion, a clumpy ISM and a preionisation phase, OCH is the same except with an off-centred explosion, while CCC and OCC represent models without a 
preionisation phase. For example, the M65OCC model is a halo with $M_{\rm{vir}} = 10^{6.5}~$M$_\odot$, which experiences an off-centred explosion (defined as a supernova at the radius enclosing half the gas mass within the scale radius) in a clumpy ISM, with no preionisation phase. Only the \citet{hawthorn15} models with 
a clumpy ISM and radiative cooling are used, as these are the most realistic.

\section{Modelled DLAs}
\label{s:comparison}

\begin{figure}
	\centering
	\includegraphics[width=.45\textwidth]{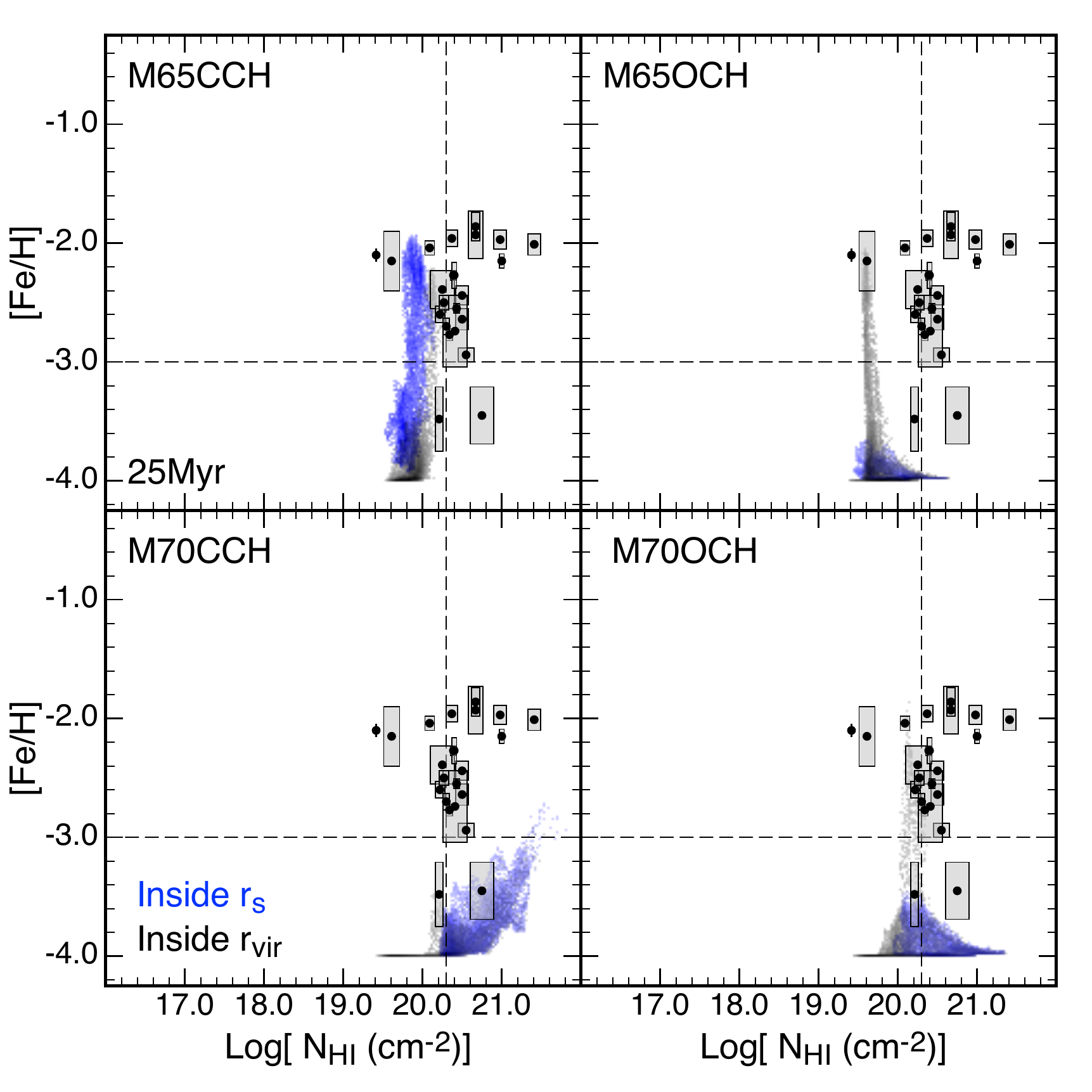}
	\caption{The distribution of H$_I$ column density vs projected metallicity 25~Myr after the explosion of a 25~M$_\odot$ star for the models with a preionization phase. The vertical dashed line indicates the definition of the minimum DLA column density. The \citet{cooke14} DLAs are also plotted with 
their error boxes.}
	\label{f:CHDLAplots}
\end{figure}

\begin{figure}
	\centering
	\includegraphics[width=.45\textwidth]{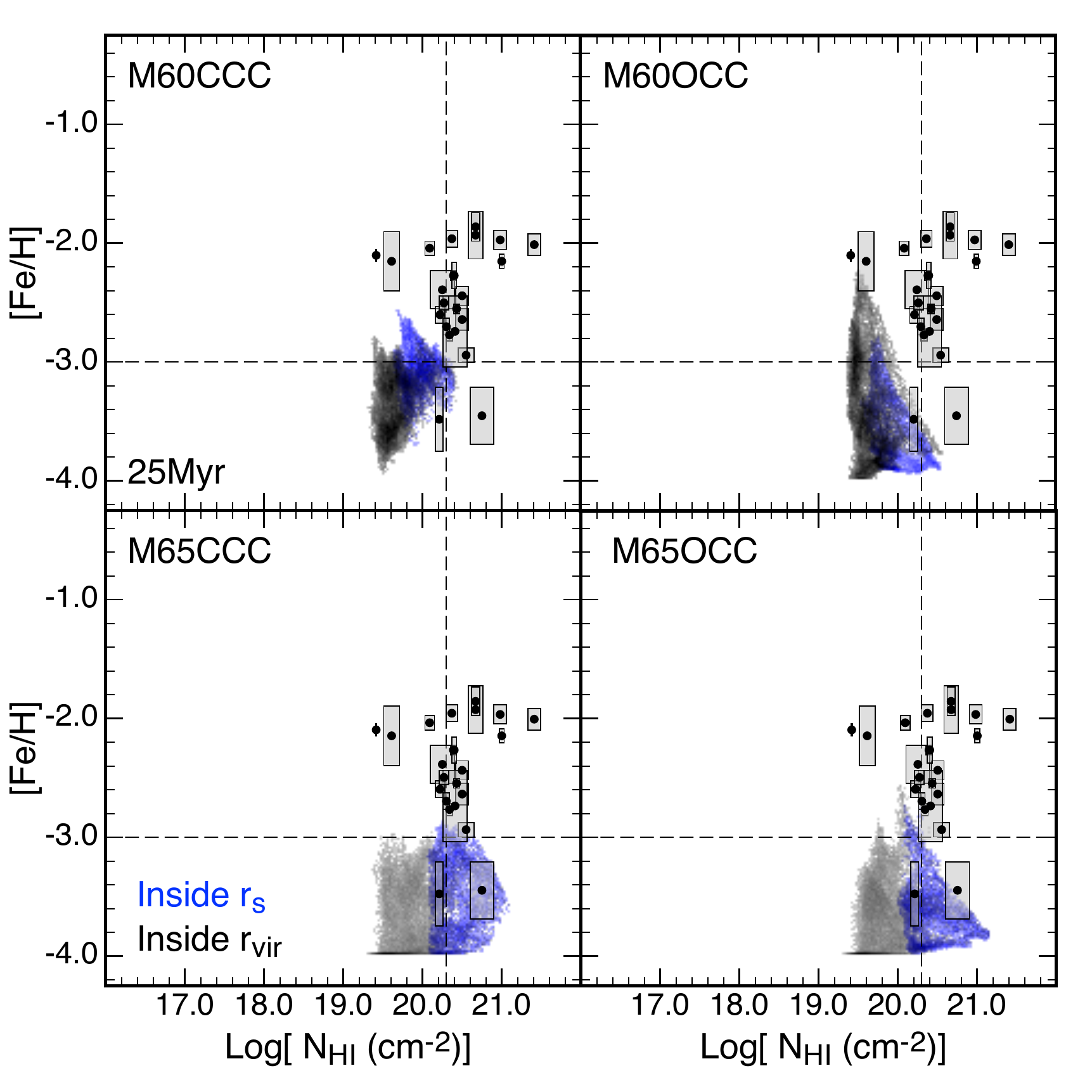}
	\caption{The distribution of H$_I$ column density vs projected metallicity 25~Myr after the explosion of a 25~M$_\odot$ star for the models without a preionization phase. The vertical dashed line indicates the definition of the minimum DLA column density. The \citet{cooke14} DLAs are also plotted with 
their error boxes.}
	\label{f:CCDLAplots}
\end{figure}

\begin{figure}
	\centering
	\includegraphics[width=.45\textwidth]{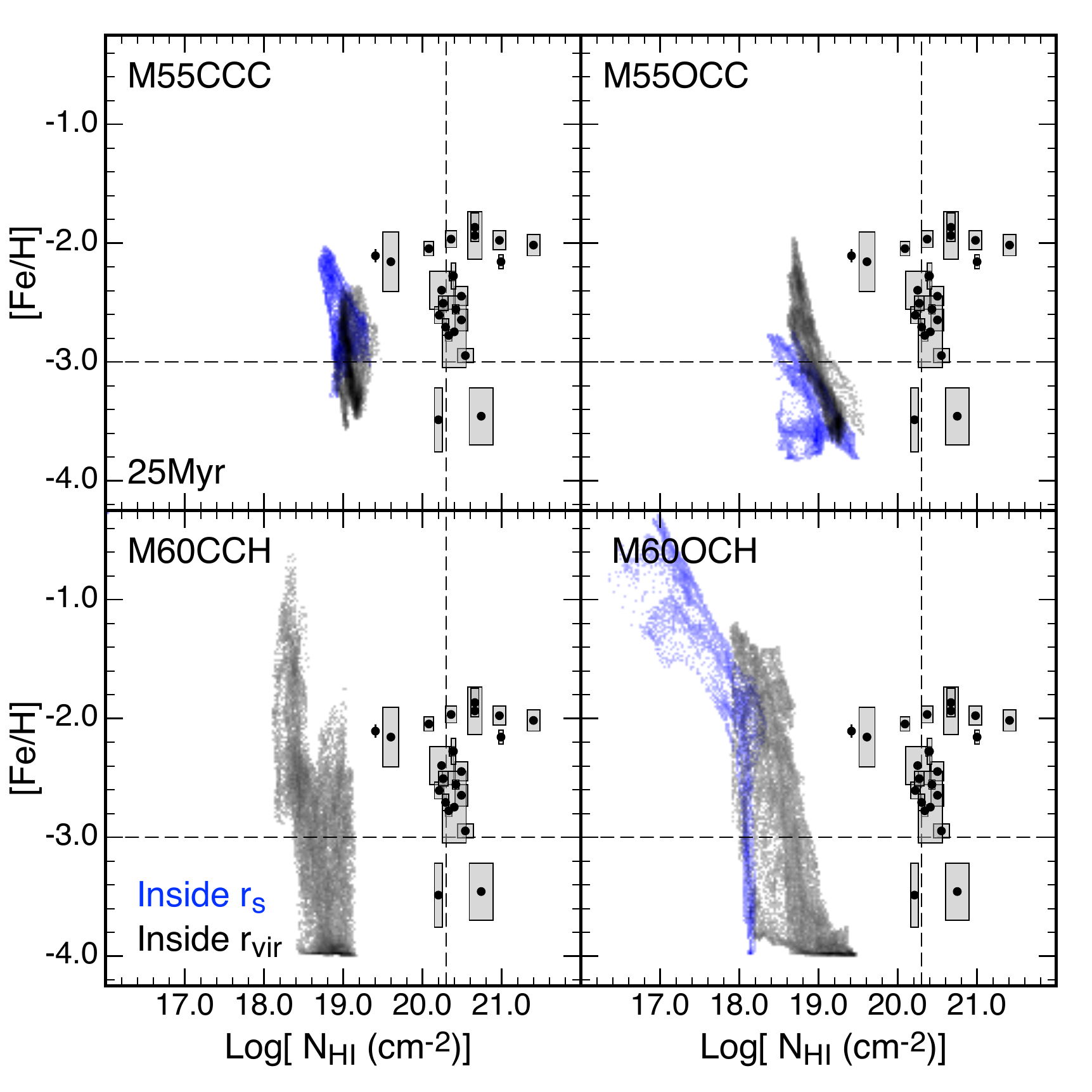}
	\caption{The distribution of H$_I$ column density vs projected metallicity 25~Myr after the explosion of a 25~M$_\odot$ star for M60 with a 
preionisation phase and M55 without a preionisation phase. The vertical dashed line indicates the definition of the minimum DLA column density. The \citet{cooke14} DLAs are also plotted with 
their error boxes.}
	\label{f:failDLAplots}
\end{figure}

In this section, we discuss the projected DLAs from the simulations discussed briefly in Section~\ref{s:simulations} and in detail in \citet{hawthorn15}. 
The column densities and metallicities result from enrichment by a single supernova from [Fe/H]~$=-4$. The metal-rich supernova ejecta were traced with 
a scalar variable, advected passively, such that the local enrichment of the gas could be determined at each time-step. Figs.~\ref{f:CHDLAplots}-\ref{f:failDLAplots} show the distribution of column densities and metallicities along projected lines of sight, while 
Figs.~\ref{f:M70CH}-\ref{f:M60CC} show spatial maps of column density and metallicity. The state of the gas after 25~Myr is used, as this is sufficient 
time for the gas to recover from the supernova \citep{webster14}. 

In this section, we assume that our halos exist in isolation and do not consider the complex possibility of mergers between low-mass gas-rich halos. A merger could introduce a fresh supply of low-metallicity gas, resulting in dilution as in the scenario considered in Section 4.2. This 
would result in higher column densities and lower metallicities, such that more sightlines would fit the DLA threshold. This would make it less likely that 
the systems could reach the metallicities of the DLAs with only a single supernova. The merger could also induce a burst of star 
formation. The effect of this is less certain, with possibilities including little or no change to the results, a more rapid increase in metallicity, 
or the removal of a large proportion of the gas in the system due to multiple supernovae close together in time.

We first discuss the models with a strong preionisation phase from a 25~M$_\odot$ star, then 
the models with a weak preionisation phase, consistent with an $M<15$~M$_\odot$ star.

\subsection{Strong preionisation phase}

\begin{figure}%[htp]
     \centering
     \includegraphics[width=.45\textwidth]{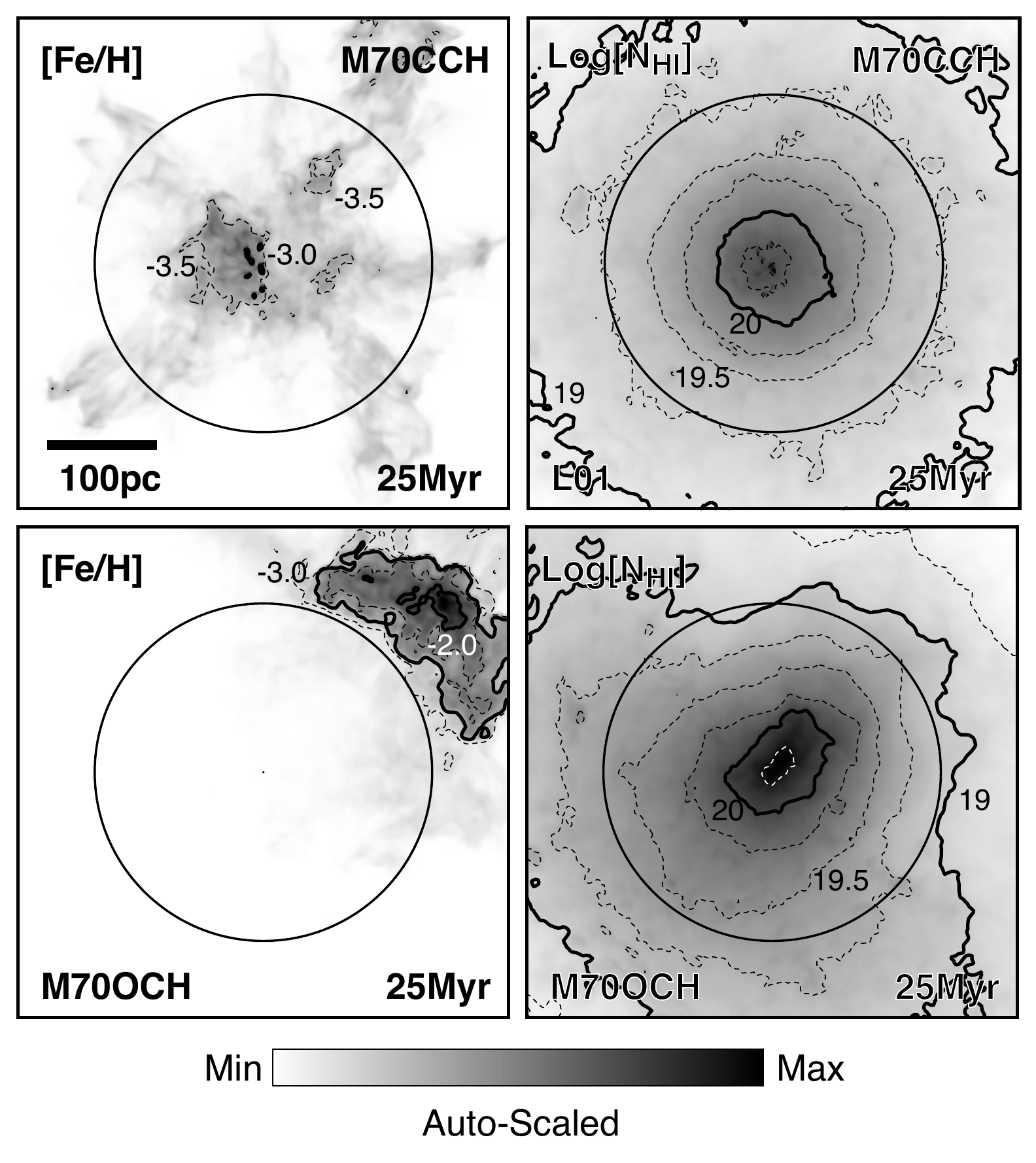}
     \caption{M70 model 25~Myr after the explosion of a 25~M$_\odot$ star. Top: Central explosion. Bottom: Off-centred explosion. The projected 
column densities are shown in the contour maps on the right, while the metallicities are shown on the left.}
     \label{f:M70CH}           
\end{figure}

As can be seen in the top panel of Fig.~\ref{f:M70CH}, the M70CCH model retains a large amount of dense gas within the scale radius (150~pc), 
with the entire region having a column density 
$N_{H_I}>10^{19.25}$cm$^{-2}$. A number of sightlines within 50~pc of the centre are dense enough to be classified as DLAs. The metallicity of 
the densest region is [Fe/H]~$\sim-3.5$, with some gas reaching [Fe/H]~$=-3.0$. The bottom-left panel of Fig.~\ref{f:CHDLAplots} shows that 
the highest density regions are the most metal-rich. This results from the central location of the supernova, such that the metals are initially 
deposited into the dense central region, along with it being the most massive model, such that more metals are retained close to the 
centre rather than reaching the halo or escaping into the IGM. No other model shows this trend. Several sightlines in M70CCH reach column densities 
greater than 
$10^{21}$~cm$^{-2}$

The M70OCH model is shown in the bottom panel of Fig.~\ref{f:M70CH}. Unlike in the central case, the dense regions do not coincide with the 
higher metallicity regions. The metals from the off-centred explosion do not reach the central region and the densest gas is therefore 
not enriched. The bottom-right panel of Fig.~\ref{f:CHDLAplots} show that there are a few sightlines outside the scale radius that 
reach [Fe/H] = $-2.5$ with column densities close to the minimum required to be defined 
as a DLA. These metallicities are higher than is seen in the central explosion model, but this is because there is less hydrogen gas in this region. The metal escape fraction is actually higher than in the central case because the metals can escape through the lower 
density regions away from the centre. Most of the dense gas has [Fe/H] $<-3.5$ and the sightlines with column densities greater 
than $10^{21}~$cm$^{-2}$ show almost no enrichment. 

\begin{figure}
          
          \includegraphics[width=.45\textwidth]{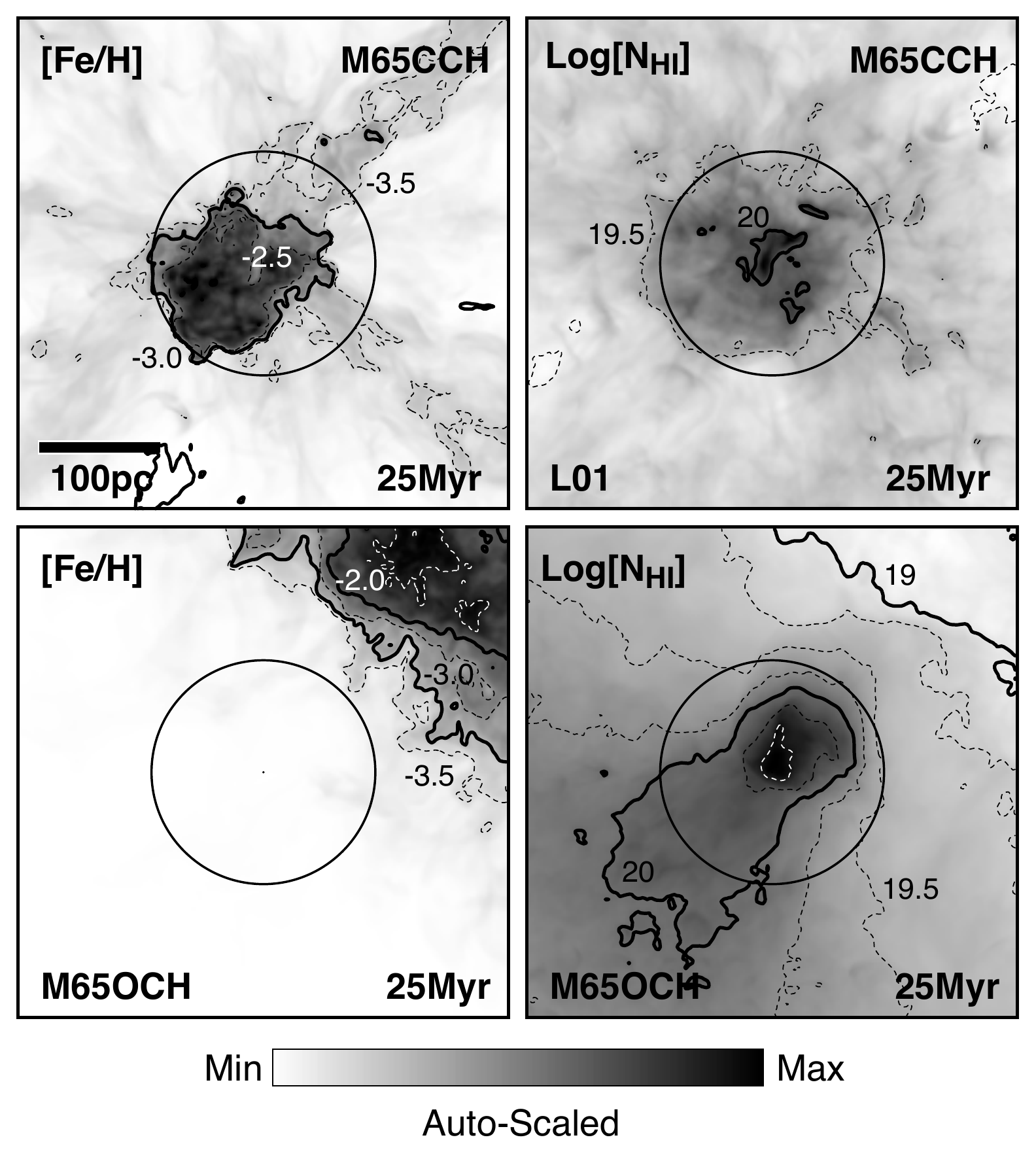}
      \caption{M65 model 25~Myr after the explosion of a 25~M$_\odot$ star. Top: Central explosion. Bottom: Off-centred explosion. The projected 
column densities are shown in the contour maps on the right, while the metallicities are shown on the left.}
	\label{f:M65CH}
\end{figure}

The M65CCH model shown in the top panel of Fig.~\ref{f:M65CH} has lower column densities than M70CCH with no sightline meeting the definition 
of a DLA. This model reaches higher [Fe/H] because there is less hydrogen. The total amount of metals is less than in the M70 case, as a greater 
proportion of the metals escape into the IGM as a result of the lower halo potential. The metallicity distribution in the top-left panel of 
Fig.~\ref{f:CHDLAplots} does not show an increase in 
metallicity with increasing column density because the enriched regions are those most affected by the supernova and therefore 
lose a significant amount of their dense gas.

The M65OCH model in the bottom panel of Fig.~\ref{f:M65CH} and the top-right panel of Fig.~\ref{f:CHDLAplots} shows similar features to the M70OCH model, although the column densities are lower and the metallicities are higher for the same reasons as discussed for M65CCH. A few sightlines exceed the DLA 
threshold, however none of these are enriched above the starting metallicity [Fe/H]~$=~-4$. The lines of sight with the highest metallicity 
[Fe/H]~$\sim-2$ have column densities of $10^{19.6-19.8}$~cm$^{-2}$.

\subsection{Weak preionisation phase}

\begin{figure}
          \includegraphics[width=.45\textwidth]{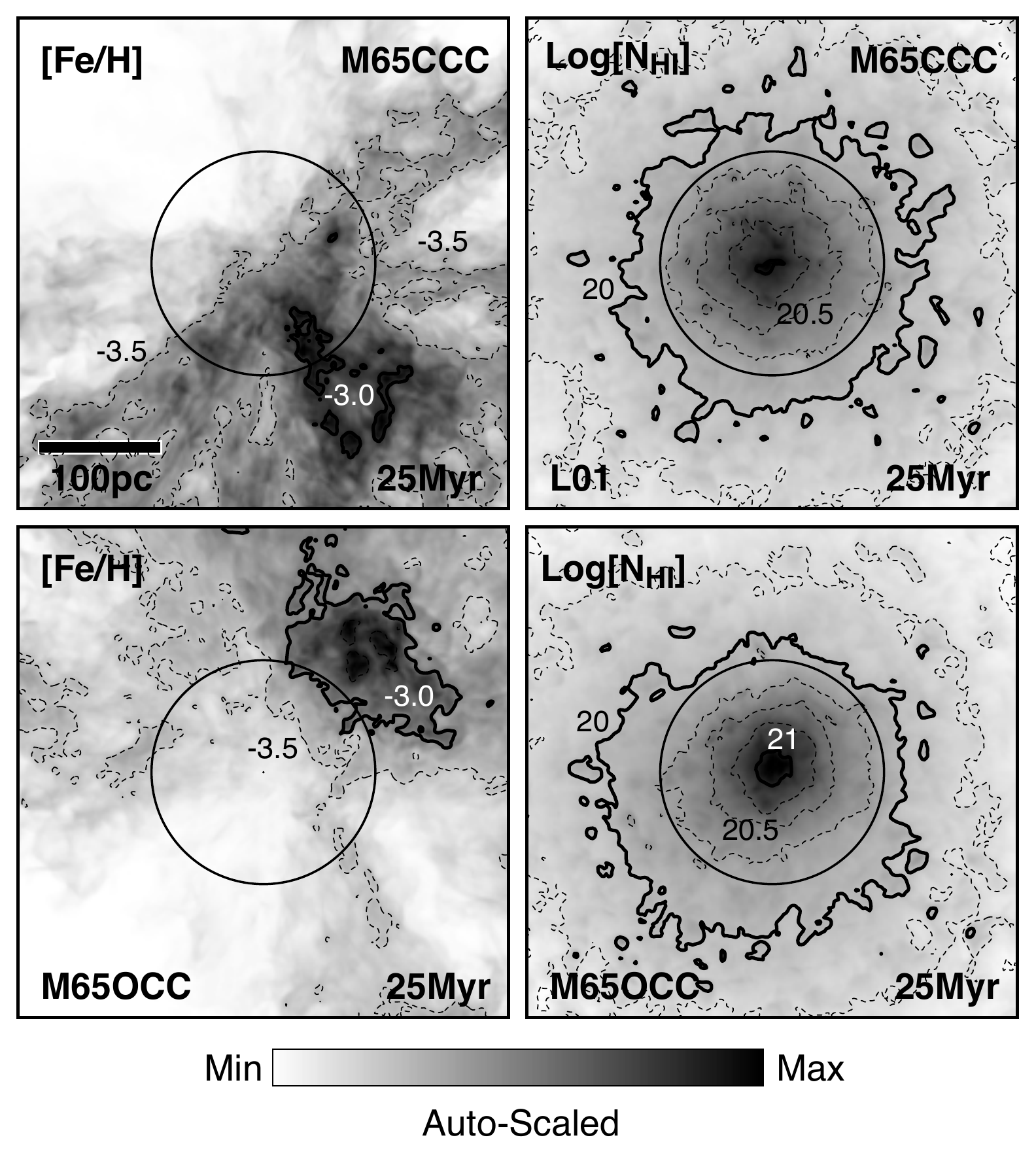}
      \caption{M65 model 25~Myr after the explosion of a star without a preionization phase. Top: Central explosion. Bottom: Off-centred explosion. The projected column densities are shown in the contour maps on the right, while the metallicities are shown on the left.}
	 \label{f:M65CC}
\end{figure}

The M65CCC model is shown in the top panel of Fig.~\ref{f:M65CC}. The column density distribution is similar to M70CCH model, which is consistent with our 
finding in \citet{hawthorn15} that a decrease of 0.5~dex in halo mass has a similar effect on gas retention as switching off the preionisation phase.  
[Fe/H] is higher than in the M70CCH model because the gas is less dense while the chemical yields from the star are the same. The metallicity distribution 
in the bottom-left panel of Fig.~\ref{f:CCDLAplots} shows almost no variation with column density.

The M65OCC model in the bottom panel of Fig.~\ref{f:M65CC} also shows a similar column density distribution to its M70 counterpart with preionisation, 
but once again the metallicity distribution is different. The lower gas densities in the M65 halo allow some metals to reach and enrich the dense 
gas in the centre. 
This can be seen in the bottom-right panel of Fig.~\ref{f:CCDLAplots}, where some of the gas within the scale radius reaches [Fe/H]~$>-3$. 

\begin{figure}
          \includegraphics[width=.45\textwidth]{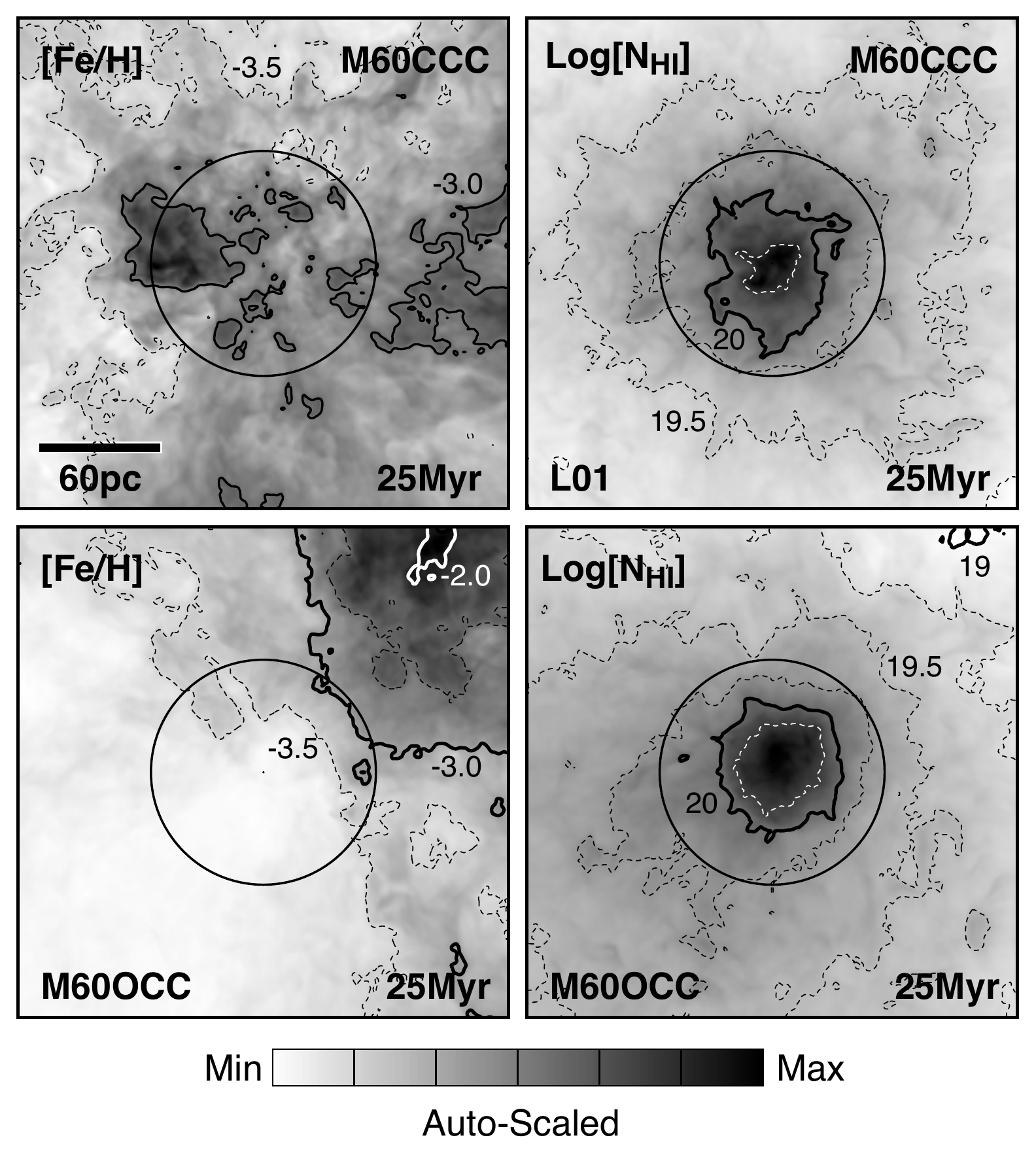}
      \caption{M65 model 25~Myr after the explosion of a star without a preionization phase. Top: Central explosion. Bottom: Off-centred explosion. The projected column densities are shown in the contour maps on the right, while the metallicities are shown on the left.}
	\label{f:M60CC}
\end{figure}

The lack of a preionisation phase means that even M60 halos can retain sufficient gas to reach the DLA threshold, as is shown in Fig.~\ref{f:M60CC}. 
In M60CCC, a few sightlines reach column densities of $10^{20.3}$~cm$^{-2}$ with [Fe/H]$~=~-3.0$. Unlike in the M65 case, there is no gas within the scale radius with [Fe/H]~$<-3.5$, showing that the supernova has enriched all the central gas. As shown in the top-left panel of Fig.~\ref{f:CCDLAplots}, 
this is the only system other than M70CCH for which metallicity increases with column density.  

The M60OCC model shows a low density, high metallicity region near the supernova which is a feature of all the models with an off-centred explosion. 
The top-right panel of Fig.~\ref{f:CCDLAplots} shows that the regions with column densities exceeding the DLA threshold remain unenriched.    

Fig.~\ref{f:failDLAplots} shows the lowest mass systems we modelled. The M60CH models and the M55CC models do not retain dense gas in the face of the 
supernova explosion, with the maximum column density being $10^{19-19.5}$~cm$^{-2}$. In the off-centred cases the densest gas remains unenriched, while 
the models with a central explosion retain gas with a column density of $10^{19}$~cm$^{-2}$ with [Fe/H]~$\sim~-2.5$. 

\subsection{Comparison with the Cooke DLAs}

In this section we compare our models above to the 23 observed very metal-poor DLAs presented by \citet{cooke14}. These DLAs have redshifts of 
$z~=~2.1-4.5$, [Fe/H] ranging from $-3.48$ to $-1.86$, and column densities $\log({N_{H_I}}/$cm$^{-2})$ = 19.6-21.4. Figs.~\ref{f:CHDLAplots}-\ref{f:failDLAplots} show that
most of the DLAs have metallicities higher than for our modelled DLAs, with the highest metallicities in our models resulting from low amounts of 
neutral hydrogen and therefore low column densities. This suggests that enrichment from a single supernova is generally not 
sufficient to explain the observed DLAs. We will discuss extended star formation in Section~\ref{s:scenarios}. However, a few DLAs could be explained 
by enrichment from a single star.

Two systems show [Fe/H] $\sim-3.5$. 
The gas masses are not known for these systems, although one has an upper limit of 
$M_{\rm{WNM}} < 6.3\times10^6$~M$_\odot$. In our M70CH models, which have a gas mass
of $10^6$~M$_\odot$ within $r_{vir}$ and $2\times10^5$~M$_\odot$ within $r_s$, the 25~M$_\odot$ star enriches the surrounding gas to metallicities 
ranging from [Fe/H] = $-4.0$ to $-3.0$. The two lowest metallicity DLAs are therefore consistent with enrichment by a single supernova in a $10^7$~M$_\odot$ 
halo. The M65CCC model also shows sightlines with the correct column density and metallicity. These two DLAs may therefore trace gas polluted only by 
a single supernova. This level of [Fe/H] is consistent with that 
suggested by \citet{bromm11} for the first galaxies, so they may even only be enriched by Population III stars.

Two other DLAs could be explained by enrichment from a single supernova if they are lower mass systems. HS0105+1619 and J2155+1358 have 
column densities $\approx10^{19.5}$~cm$^{-2}$ and [Fe/H] $\approx-2.1$. This is consistent with the M65CH models. However, both show 
[$\alpha$/Fe] $< 0.3$, which is difficult to explain given enrichment from a 25~M$_\odot$ star. The chemical abundances are more consistent with the 
higher column density systems, so it is more likely that they have experienced an extended period of star formation followed by losing most of their gas. The 
column densities are consistent with the systems that form no further stars in our models.

The remainder of the DLAs have metallicities too high to be explained by a single supernova. However, in \citet{webster14} we showed that 
the gas recovers from the impact of a 25~M$_\odot$ star in less than 25~Myr for an M70 halo regardless 
of supernova location. Furthermore, the M65 models 
without a preionisation phase show little evidence of disruption or gas loss, suggesting that stars less massive than 25~M$_\odot$ are unlikely to 
blow out a significant proportion of the gas in an M70 system. Our DLA models are therefore likely to remain relevant for subsequent supernovae.

\begin{table}
\caption{Properties of DLAs compared to our model. The masses, radii and densities for the models are taken at the scale radius of the Einasto potential.}
\begin{tabular}{c c c c c c}
&M70&M65&$<\rm{DLAs}>$&DLAs$_{\rm{max}}$&DLAs$_{\rm{min}}$\\
\hline
$T_{gas}$ (K)&4860&2300&9600&17000&5600\\
log$(n(H)/$cm$^{-3})$&-0.13&-0.07&-1.0&-0.35&-1.3\\
$r_{H_{\rm{I}}}$ (pc)&150&91&220&1270&32\\
$M_{\rm{WNM}}$ (10$^5$M$_\odot$) &2.3&0.7&2.5&220&$\leq0.4$\\
$c_s$&5.8&4.0&8.0&10.7&6.1\\
\hline
\\
\end{tabular}
\end{table}

\citet{cooke14} determined the thermal contribution to the line broadening for 12 clouds in 9 of the 23 systems, allowing gas properties to be calculated.
The total warm neutral medium gas masses range from less than 
$3\times10^4$ to $2\times10^7$~M$_\odot$. A comparison between the properties of the gas in our model and the properties 
determined for these DLAs is shown in Table 1. The neutral hydrogen density $n(H)$, the cloud radius $r_{H_{\rm{I}}}$ and the warm neutral gas mass 
$M_{\rm{WNM}}$ for our model are at the scale radius. It is possible that the gas observed in DLAs 
extends beyond these radii, which would result in lower densities and higher gas masses. The temperature and sound speed are slightly lower in the models 
than for the observations. Overall, the DLAs for which these physical quantities could be determined are consistent with halos with $10^7$~M$_\odot$ or slightly higher. 

\section{Enrichment of the DLAs}

In this section we will discuss two possible enrichment histories for the DLAs that show evidence of extended star formation. 
The first scenario suggests that they have been enriched only by a single burst of stars, with the alternative being that they are a mix of 
systems which have been enriched by one burst and those which have been enriched by two bursts following the accretion of metal-poor gas. As discussed 
in \citet{webster14} \& \citet{hawthorn15}, 
only the M70CH and the M65CC models retain dense enough gas for extended low mass star formation. This section focusses on the M70CH model, using the 
chemical evolution model of \citet{webster14}.

\label{s:scenarios}
\subsection{Single burst}

The single burst scenario explains the relationship between [$\alpha$/Fe] and [Fe/H] in the following way: 
\begin{enumerate}
\item{Gas enriched to [Fe/H] $\sim$ -4 condenses onto a dark matter halo.}
\item{Low mass star formation commences and Type II supernovae enriches the gas, resulting in an increase in [Fe/H] and a decrease in [$\alpha$/Fe]. 
[$\alpha$/Fe] declines at low [Fe/H] due to Type II supernovae from lower mass (8-15~M$_\odot$) stars, which have a mean [$\alpha$/Fe] of 0.2 \citep[][ extrapolated to 8~M$_\odot$]{woosley95,nomoto06}. This decline commences at lower [Fe/H] 
than in our model.}
\item{After $t\sim100$~Myr, Type Ia supernovae enrich the gas, resulting in a decline in [$\alpha$/Fe] with increasing [Fe/H]. This decline is observed in 
DLAs with [Fe/H]~$>-2.0$ \citep{vladilo11}.}
\end{enumerate}

\begin{figure}[htp]             
      \includegraphics[width=.45\textwidth]{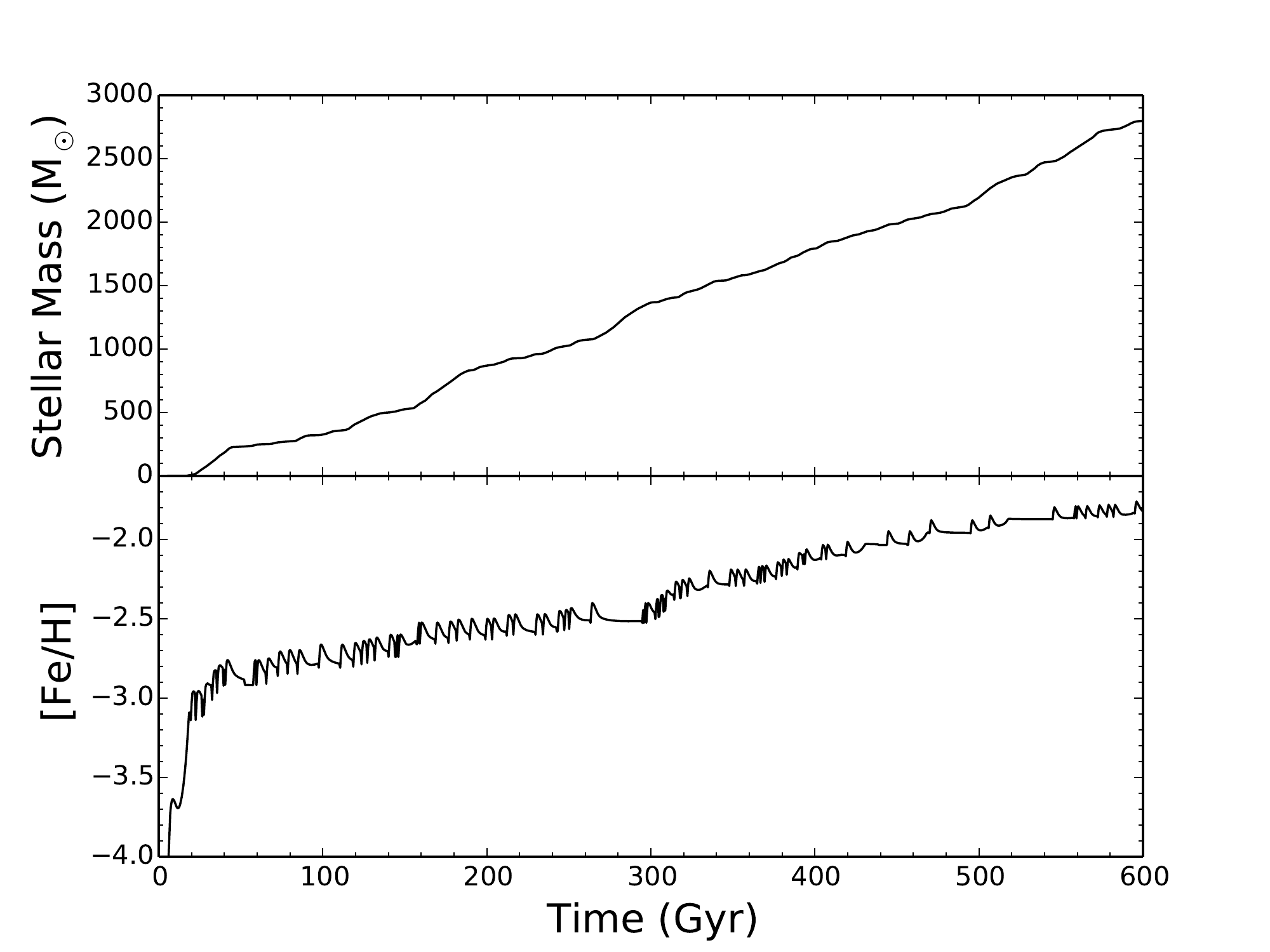}
      \caption{The gas and metallicity evolution as a function of time for one realisation of the single-burst scenario in our simulations. Top: The 
total star formation since $t=0$. Bottom: The mean metallicity of the gas within the ($200$~pc)$^3$ main simulation level as a function of time.}
\label{f:tevol}
\end{figure}

The stellar mass and gas metallicity evolution in one simulation run for this scenario is shown in Figure~\ref{f:tevol}. Our model is stochastic, such that 
different simulation runs will provide different results, but the overall features are similar. The star formation rate is on average constant with time, 
but is bursty, with several short periods of little or no star formation following each supernova explosion. Each supernova causes a large 
jump in the mean metallicity in the central region, which is caused by the deposition of new metals onto the grid, along with the lower metallicity gas being pushed outwards beyond the inner (200~pc)$^3$ region considered here. The mean metallicity in this region then decreases as gas returns to the centre, 
although some of the metals have mixed with the gas and the overall metallicity therefore remains higher than before the supernova.

\begin{figure*}[htp]
     \centering               
      \includegraphics[width=.9\textwidth]{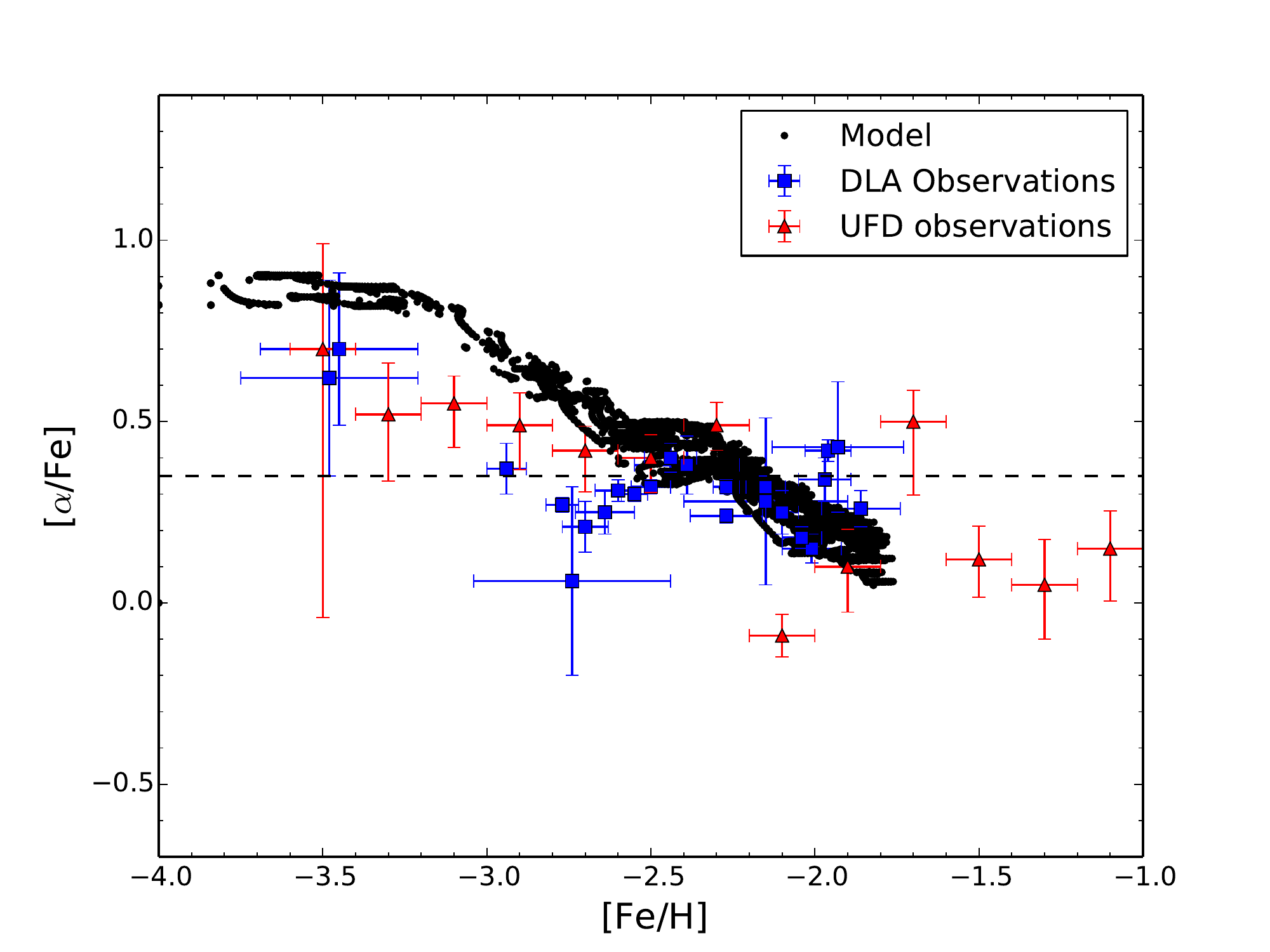}
      \caption{[$\alpha$/Fe] vs [Fe/H] for our simulation for five simulation runs (black points), the \citet{vargas13} sample of stars in UFDs in 0.2~dex metallicity bins (red triangles), and the \citet{cooke14} DLAs (blue squares).}
\label{f:UFDafeplot}
\end{figure*}

Figure~\ref{f:UFDafeplot} shows [$\alpha$/Fe] vs [Fe/H] for the \citet{cooke14} DLA observations compared to the gas in 
our simulation, as well as observations of stars in UFDs from \citet{vargas13}. The three are reasonably consistent for [Fe/H]~$\gtrsim -2.5$. However, 
at lower metallicities both our model and the UFD observations show a gradual decline in [$\alpha$/Fe] 
with [Fe/H], while the DLAs show a rapid decline between [Fe/H] $\sim$ $-3.5$ and $-2.8$, with [$\alpha$/Fe] then remaining constant until [Fe/H] = $-2.0$.
The mean [$\alpha$/Fe] abundance of 0.25 for the DLAs for $-3<$~[Fe/H]$<-2.5$ is also lower than the average of 0.35 
for Type II supernovae in a Kroupa or Salpeter IMF, which is observed in stars in the halo of the Milky Way \citep{frebel12}.

\citet{cooke14} also notes the suppression in [$\alpha$/Fe] for the DLAs compared to stars in the overall population of dSph galaxies. 
They suggest that this could be caused by 
small number statistics, modelling techniques, or physical differences between the two populations. \citet{karlsson12} suggested two classes 
of explanations for suppressed [$\alpha$/Fe] in a star cluster in Sextans. 
The first explains the low [$\alpha$/Fe] by contribution from Type Ia supernovae at low [Fe/H]. This could result either from the accretion of low metallicity gas, reducing [Fe/H] at a time when Type Ia supernovae contribute to the enrichment, or from a very low star formation rate. The second 
class of explanations involve only Type II supernovae. \citet{aoki09} note that a truncated IMF at high masses would result in 
lower [$\alpha$/Fe], because higher mass stars yield more alpha elements. \citet{weidner05} suggest that such an IMF is required for systems with 
very low star formation, which are likely to have low cluster masses, limiting the maximum mass for a star that can form in the cluster. The final 
alternative discussed is hypernovae, which produce more Fe than normal Type II supernovae. However, based on iron-peak element abundances, 
\citet{cooke13} suggest that very metal-poor DLAs were enriched by stars that exploded as core collapse supernovae which 
released 1.2$\times10^{51}$~erg of energy.

The question of whether Type Ia supernovae are responsible for the suppression of [$\alpha$/Fe] in DLAs is not easily resolved. 
However, assuming the DLAs are first 
burst systems and a reasonably homogeneous population, 
explanations for the decline in [$\alpha$/Fe] between [Fe/H] = $-3.5$ and $-2.5$ 
involving Type Ia supernovae are unable explain the constant [$\alpha$/Fe] for $-2.5<$~[Fe/H]~$<-2.0$. It is more likely that for a single burst, 
Type Ia supernovae start occurring at [Fe/H]~$\approx-2.0$ as was concluded by \citet{cooke14}. 

\begin{figure*}%[htp]
     \centering       
          \includegraphics[width=.9\textwidth]{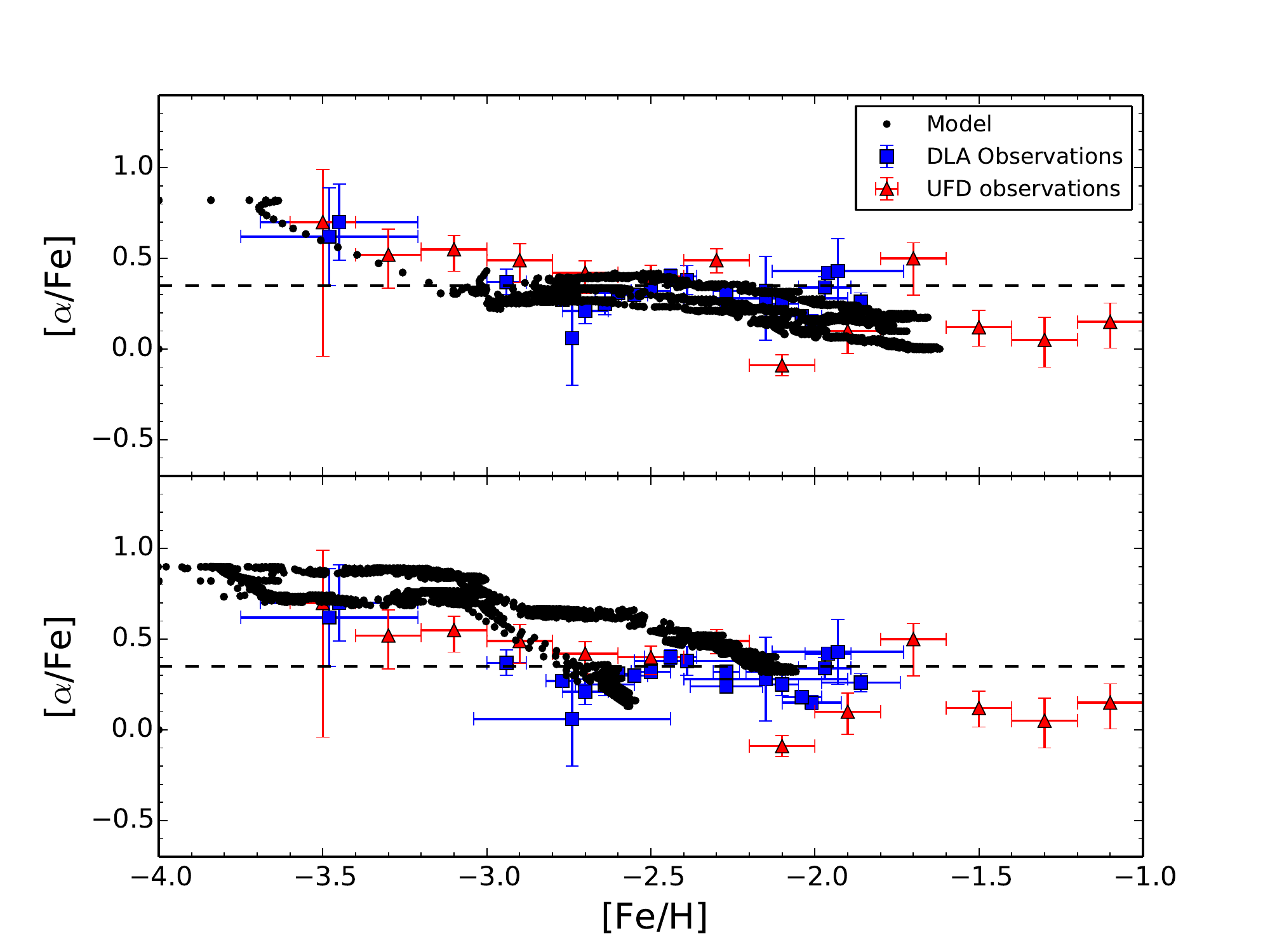}
      \caption{Top: Single burst scenario with truncated IMF at the high mass end. Bottom: Two burst 
scenario with accretion of low [Fe/H] gas after the first burst. [$\alpha$/Fe] vs [Fe/H] for three simulation runs in each plot is represented by black points, the \citet{vargas13} sample of stars in UFDs in 0.2~dex metallicity bins by red triangles, and the \citet{cooke14} DLAs by blue squares.}
\label{f:DLAafeplotsscen}    
\end{figure*}

Given that enrichment by Type Ia supernovae is unlikely to explain the low [$\alpha$/Fe] for $-2.5<$~[Fe/H]~$<~2.0$, the single burst scenario 
requires that supernovae with lower mass yields have enriched the gas observed in DLAs. One possibility for this is that DLAs are enriched by low-mass 
($8-15$~M$_\odot$) Type II supernovae to a level lower than the average [$\alpha$/Fe] for a Kroupa 
or Salpeter IMF. One possible explanation for this is that for the reasons discussed above, the 
IMF is truncated at the high mass end. We model an IMF truncated at 20~M$_\odot$, with 
the results shown in the top panel of Fig.~\ref{f:DLAafeplotsscen}. [$\alpha$/Fe] remains slightly enhanced compared to the [Fe/H]~$\sim-2.5$ DLAs. 
Reducing the highest mass further would fit the lowest metallicity DLAs, but could 
not simultaneously explain the DLAs with [Fe/H]~$>-2.5$.
  
Another possible explanation is a selection effect. In \citet{hawthorn15}  
we studied the effect of a 25~M$_\odot$ star on halos with masses $10^6-10^7$~M$_\odot$. It was found that a strong preionization phase evacuates 
the region close to the star, resulting in the subsequent supernova having a much larger effect on the ISM than it would in the case of a $<15$~M$_\odot$ 
star with a much weaker preionization phase. In particular, a 25~M$_\odot$ star in a $10^{6.5}$~M$_\odot$ halo can terminate star 
formation \citep{webster14}.
 It is therefore possible that systems which have experienced lower mass stars and therefore 
lower [$\alpha$/Fe] are more likely to contain dense gas and be observed as DLAs. This would have a similar effect to the truncated IMF as in the top 
panel of Figure~\ref{f:DLAafeplotsscen}. 

The above explanations provide a number of possible ways that Type II supernovae could result in the suppressed [$\alpha$/Fe] seen in very metal-poor DLAs compared to the Kroupa IMF. However, it does not explain why the DLAs show lower [$\alpha$/Fe] compared to stars in UFDs, showing [$\alpha$/Fe] lower by 0.25~dex for $-3.0<$~[Fe/H]~$<-2.5$.
 None of the observed UFDs has mean [$\alpha$/Fe] $<$ 0.4 for [Fe/H] $<-2.5$. We therefore conclude that if all DLAs are systems forming their 
first burst of stars, either they are not representative of the UFD population, or there are systematic differences that arise in the process of 
measuring [$\alpha$/Fe] between gas in DLAs and stars in UFDs. It is possible that as a result of evolution, the DLAs are not representative of 
UFDs as observed today. Many of the DLAs contain sufficiently dense gas to continue forming stars. In the next section we consider a scenario 
where the difference between the two classes of systems is explained by the DLAs being observed at different points in their evolution.

\subsection{Two bursts}

We now investigate a scenario in which the DLAs are a mix of systems that have formed only one burst of stars, and systems that have 
formed or are forming their second burst of stars after the accretion of metal-poor gas. 
\citet{weisz14b} investigated the star formation history of 38 Local Group dwarf galaxies, 
including 4 of the UFDs studied in \citet{vargas13}. They 
found that Hercules was consistent with having formed all its stars in one burst before and during reionization. Leo IV also shows only one burst, 
but likely formed stars until $z\sim2$. Leo T and Canes Venatici II have a more extended star formation history, 
although Leo T shows no star formation from $z~=1-5$. The best fit model to Canes Venatici II shows a burst of star formation commencing at $z\sim2.5$, 
while the 1$\sigma$ uncertainty is consistent with 80\% of the stars being formed in a single burst ending at $z\sim2.5$. This suggests that
 star formation occurred in some UFDs at the redshifts of the \citet{cooke14} DLAs, and that some were forming their first burst of stars, while others 
were forming their second burst.

We now compare the DLA population to the stars in Leo IV and 
Canes Venatici II, the two systems which are likely to form stars at the redshifts of the \citet{cooke14} DLAs. The results are shown in the bottom 
panel of  Fig.~\ref{f:DLAafeplotsscen}. The 4 stars in Leo IV are consistent with the [$\alpha$/Fe] abundances of the DLAs, while of the 
6 stars in Canes Venatici II 
with [Fe/H]$<-1.5$, 4 show [$\alpha$/Fe] enhanced significantly above that of the DLAs. It should be noted that while \citet{weisz14b} find that 
the best fit for the star forming history of Canes Venatici II 
is a model with multiple bursts, \citet{brown14} argue for a single burst. We present the following possible scenario for its evolution. 
The stars with [$\alpha$/Fe]~$\sim0.8$ formed in the first burst, followed by reionization or supernova feedback turning star formation off. 
The galaxy accreted low-metallicity gas which was then polluted by Type Ia supernovae. The second burst of stars therefore 
formed at a lower [$\alpha$/Fe] for a given [Fe/H]. 

If the DLAs trace gas in UFDs, it is likely that we are observing some of them during or before their second burst of star 
formation. This provides an explanation as to why the DLAs tend to have lower [$\alpha$/Fe] at low [Fe/H] when compared to the UFDs 
and the average expected from Type II supernovae. High [$\alpha$/Fe] stars formed in the first burst of star formation before and during reionization, followed by a pause in star formation. By the 
time of the second burst of star formation, sufficient time has passed for Type Ia supernovae to occur, resulting in the reduced [$\alpha$/Fe] seen in 
the DLAs. This also provides an explanation for low [$\alpha$/Fe], low [Fe/H] stars in systems such as Canes Venatici II and Ursa Major I.

In the two-burst model, the DLAs in the clump at [Fe/H] $\sim-2.0$ in Fig.~\ref{f:DLAafeplotsscen} are forming their first burst of stars, 
while those in the clump at [Fe/H] $\sim-2.5$ are forming their second burst. This scenario predicts the existence of [$\alpha$/Fe]~=~0.5-0.7 
systems at [Fe/H]~$\sim-2.5$, 
which have not yet been observed. 

\subsection{Other possibilities}

A third possibility is that the DLAs do not evolve from the clump at [Fe/H]~$\sim-2.5$ to the clump at [Fe/H] $\sim-2.0$, but that 
they instead represent systems with different masses and star formation histories. \citet{cooke14} found that systems with higher velocity widths showed higher metallicities. While it is impossible to extrapolate directly to halo mass, systems with lower velocity widths will on average have lower 
halo masses. The [Fe/H] $\sim$ -2.5 systems may therefore represent systems which correspond to UFDs, while the higher metallicity systems may be classical 
dwarf spheroidals. However, this does explain why the DLAs are observed with [$\alpha$/Fe] $\sim$ 0.3, when Type Ia supernovae could suppress [$\alpha$/Fe] below this level. 

Our models in Section~\ref{s:scenarios} showed that systems such as 
M65CH which do not show star formation can still reach the DLA column density limit along some sightlines. Some of the DLAs may therefore be systems that 
are not forming stars at the time they are observed. The clump at [Fe/H]~$\sim-2.5$ could be systems which formed few stars, with most 
of the enrichment coming from later Type Ia supernovae, while the [Fe/H]~$\sim-2.0$ systems can be explained by the single-burst model. Observations of 
neutron-capture elements such as barium may be able to provide support for or rule out this scenario. Neutron-capture abundances could also 
test the possibility that the [$\alpha$/Fe] abundances result from a bimodal Type Ia supernova delay-time distribution \citep{mannucci06,yates13}, 
in which approximately 50\% of the supernovae occur at $t\sim50$~Myr. This could result in an early decline in [$\alpha$/Fe], followed by a period of 
few Type Ia supernovae, during which [$\alpha$/Fe] is flat with [Fe/H], as is observed in the DLAs.

An alternative explanation for the variation in [$\alpha$/Fe] is the enrichment of some DLAs by pair-instability supernovae. 
Pair-instability supernovae eject a large amount of iron, resulting in low [$\alpha$/Fe]. \citet{wise12} suggests this as an explanation for the apparent 
metallicity floor of DLAs at [Fe/H] $=-3$. 
The only exceptions are two DLAs with high [$\alpha$/Fe] show [Fe/H] $< -3$, which these may have been enriched by an event 
with more usual supernova yields. In future work we will investigate different types of supernovae, such as hypernovae and pair-instability supernovae, 
to determine their effect on our higher mass models.

Finally, we note that factors outside the scope of our model could affect our result. For example, in this work we assume that all metal species mix into the gas in the same way. While 
different elements may have different dust formation time-scales, the gas and dust physics involved is difficult to track. The best work to date on this 
issue has involved considering the dust and molecular gas phases in post-processing \citep{krumholz11,gnedin14}. These complicated issues will be 
considered in future work. However, none of the scenarios in this subsection can resolve the discrepancy between the UFDs and the DLAs, requiring different 
enrichment histories for the two classes of systems

\section{Conclusions}
\label{s:conclusions}

Using simulations of star formation and chemical enrichment in systems with dark matter halo masses of $10^7$~M$_\odot$, we have modelled the chemical 
abundances of very metal-poor DLAs. Our conclusions follow:

\begin{enumerate}
\item For the DLAs in which \citet{cooke14} were able to derive physical quantities, the measured sound speeds, temperatures, gas masses and gas densities 
 are mostly consistent with gas in our modelled 
$M_{\rm{vir}} = 10^7$~M$_\odot$ halo, although the temperature and sound speed are slightly too low, suggesting the observed DLAs may have slightly 
larger halo masses. 
\item The column densities of the \citet{cooke14} DLAs can be reproduced by models of gas in a $10^7$~M$_\odot$ halo which include feedback from 
the preionization and supernova of a massive star.
\item Multiple supernovae are required to explain the metallicities of 21 of the 23 \citet{cooke14} DLAs, except in the case of a pair-instability 
supernova as in \citet{wise12}.
\item Our model of star formation and chemical enrichment can reproduce [$\alpha$/Fe] for DLAs with $-2.5\lesssim$~[Fe/H]~$\lesssim-2.0$. While 
there is significant scatter for individual systems, the average UFD abundances also agree.
\item For $-3.0\lesssim$~[Fe/H]~$\lesssim-2.5$ there is some tension between our model and the DLA abundances, with the DLAs showing mean 
[$\alpha$/Fe] = 0.25, while our model has [$\alpha$/Fe] $\sim$ 0.5. [$\alpha$/Fe] is also suppressed for the DLAs compared to the average stellar 
metallicity in UFDs and compared to the mean expected for Type II supernovae for a Kroupa or Salpeter IMF.
\item One explanation for the abundances of DLAs is a scenario with a truncated IMF, or a selection effect where only systems with 
lower mass supernovae retain their gas and are therefore observed as DLAs. However, this does not explain the the suppression of [$\alpha$/Fe] compared to 
abundances of stars in UFDs.
\item A scenario that can explain the mismatch between the DLA and UFD abundances assumes that the [Fe/H]~$=-3.0$ to $-2.5$ DLAs have been enriched 
by two bursts, while the higher [Fe/H] DLAs 
have been enriched by only one. Two-burst DLAs form their first burst before reionization before losing all their neutral gas. At a later time, they 
accrete low-metallicity gas and commence a second burst of star formation. Type Ia supernovae from first burst stars enrich the gas without a delay time. 
However, this scenario does not explain why no DLAs have been observed 
at [Fe/H]~$\sim-2.5$ with high [$\alpha$/Fe].
\item It is possible that the [Fe/H]~$=-3.0$ to $-2.5$ and the [Fe/H]~$=-2.5$ to $-2.0$ DLAs have very different star formation histories. For example, 
the low [Fe/H] systems may have formed few stars and therefore experienced few or no Type II supernovae. 
These systems were then enriched at a later time by one or more Type Ia supernovae, resulting in the low observed [$\alpha$/Fe]. The higher [Fe/H] systems 
would then be explained by the single burst scenario. 
\end{enumerate}

\acknowledgments We thank the anonymous referee for useful comments which greatly improved the clarity and quality of this work. DW is funded by an Australian Postgraduate Award.  

\bibliographystyle{apj}
\bibliography{refs}

\end{document}